\documentclass[aps,prl,showpacs,preprintnumbers,%
               twocolumn,groupedaddress,amsfonts]{revtex4}
\usepackage{graphicx}% Include figure files
\usepackage{dcolumn}% Align table columns on decimal point
\usepackage{bm}% bold math
\usepackage{amssymb,amsmath}%
\usepackage{color}
\suppressfloats[!]

\newcommand{\Pt}{p_{T}}

\newcommand{\Ptg}{p_{T}^{\gamma}}
\newcommand{\lt}{\!<\!}
\newcommand{\gt}{\!>\!}
\newcommand{\gpj}{$p\bar{p} \rightarrow \gamma + \mathrm{jet} + X $}

\newcommand{\newscale}{p_T^\gamma f(y^\star)}
\newcommand{\Fystar}{\{[1+\exp(-2|y^\star|)]/2\}^{1/2}}
\newcommand{\ystar}{0.5(y^\gamma-y^{\text {jet}})}

\newcommand{\la}{\langle}
\newcommand{\ra}{\rangle}

\begin{document}
\preprint{FERMILAB-PUB-08/081-E}
\title{Measurement of the differential cross section for the production of an isolated photon with
associated jet in $\bf p\bar{p}$ collisions at $\bf \sqrt{s}=$1.96 TeV}
%
% LIST_OF_AUTHORS_R2.TEX               3/27/08              
%
\author{V.M.~Abazov$^{36}$}
\author{B.~Abbott$^{75}$}
\author{M.~Abolins$^{65}$}
\author{B.S.~Acharya$^{29}$}
\author{M.~Adams$^{51}$}
\author{T.~Adams$^{49}$}
\author{E.~Aguilo$^{6}$}
\author{S.H.~Ahn$^{31}$}
\author{M.~Ahsan$^{59}$}
\author{G.D.~Alexeev$^{36}$}
% ...

\affiliation{\vspace{0.1 in}(The D\O\ Collaboration)\vspace{0.1 in}}
% ...
\affiliation{$^{6}$University of Alberta, Edmonton, Alberta, Canada,
                Simon Fraser University, Burnaby, British Columbia, Canada,
                York University, Toronto, Ontario, Canada, and
                McGill University, Montreal, Quebec, Canada}
% ...
\affiliation{$^{29}$Tata Institute of Fundamental Research, Mumbai, India}
% ...
\affiliation{$^{31}$Korea Detector Laboratory, Korea University, Seoul, Korea}
% ...
\affiliation{$^{36}$Joint Institute for Nuclear Research, Dubna, Russia}
% ...
\affiliation{$^{49}$Florida State University, Tallahassee, Florida 32306, USA}
% ...
\affiliation{$^{51}$University of Illinois at Chicago,
                Chicago, Illinois 60607, USA}
% ...
\affiliation{$^{59}$Kansas State University, Manhattan, Kansas 66506, USA}
% ...
\affiliation{$^{65}$Michigan State University,
                East Lansing, Michigan 48824, USA}
% ...
\affiliation{$^{75}$University of Oklahoma, Norman, Oklahoma 73019, USA}
% ...

\begin{abstract}
The process $p\bar{p}\rightarrow \gamma + \mathrm{jet} + X $ is studied using 
1.0 fb$^{-1}$ of data collected by the D0 detector at the 
Fermilab Tevatron $p\bar{p}$ collider at a center-of-mass energy 
$\sqrt{s}=$1.96 TeV. Photons are reconstructed in the central rapidity 
region $|y^\gamma|\lt 1.0$ with transverse momenta in the range $30\lt \Ptg \lt400$ GeV
while jets are reconstructed in either the central $|y^{\mathrm{jet}}|\lt 0.8$ or 
forward $1.5\lt|y^{\mathrm{jet}}|\lt 2.5$ rapidity intervals with $p_T^{\mathrm{jet}} > 15$ GeV.
The differential cross section $\mathrm{d^3}\sigma / \mathrm{d}\Ptg\mathrm{d}y^{\gamma}\mathrm{d}y^{\mathrm{jet}} $ is measured as a function of $\Ptg$ in four regions, differing by the relative 
orientations of the photon and the jet in rapidity.
Ratios between the differential cross sections in each region are also 
presented.
Next-to-leading order QCD predictions using 
different parameterizations of parton distribution functions and 
theoretical scale choices are compared to the data. The predictions do not 
simultaneously describe the measured normalization and 
$\Ptg$ dependence of the cross section in the four measured regions.

\end{abstract}
\pacs{13.85.Qk, 12.38.Qk}
\date{April 7, 2008}
\maketitle
\label{sec:Intro}

The production of a photon with associated jets in the final state is a 
powerful probe of the dynamics of hard QCD interactions \cite{PAurLindf80,JFOwens,Cont,Au2,Vo1,Mar}. 
Different angular configurations between the photon and the jets can be used to extend 
inclusive photon production measurements 
\cite{UA2_phot,CDF_phot,D0Run1_phot,Photon_paper_erratum} 
and simultaneously test the underlying dynamics of QCD hard-scattering 
subprocesses in different regions of parton momentum fraction $x$ and 
large hard-scattering scales $Q^2$.

In this Letter, we present an analysis of photon plus jets production in $p\bar{p}$ 
collisions at a center-of-mass energy $\sqrt{s}=$1.96 TeV in 
which the most-energetic (leading) photon is produced centrally with 
a rapidity $|y^{\gamma}|<1.0$ \cite{RAP}. 
The cross section as a function of photon transverse 
momentum $\Ptg$ is measured differentially for four separate angular 
configurations of the highest $p_T$ (leading) jet and the 
leading photon rapidities. The leading jet is required to be in either the 
central ($|y^{\mathrm{jet}}|\lt 0.8$) or forward ($1.5\lt|y^{\mathrm{jet}}|\lt 2.5$) rapidity intervals, 
with $p_T^{\mathrm{jet}} > 15$ GeV, and the four 
angular configurations studied are: central jets with $y^{\gamma}\!\cdot\!y^{\mathrm{jet}}>0$ 
and with $y^{\gamma}\!\cdot\! y^{\mathrm{jet}}<0$, and forward jets with 
$y^{\gamma}\!\cdot\! y^{\mathrm{jet}}>0$ and with $y^{\gamma}\!\cdot\!y^{\mathrm{jet}}<0$.
The total $x$ and $Q^{2}$ region covered by the measurement is $0.007\lesssim x \lesssim 0.8$ and 
$900 \leq Q^2 \equiv (\Ptg)^2 \leq 1.6 \times 10^{5} ~\rm GeV^2 $, 
extending the kinematic reach of previous photon plus jet 
measurements \cite{ISR,UA2_g,CDF2,H1_gp,H1_g,ZEUS_g,ZEUS_gj}. Ratios between the differential 
cross sections in the four studied angular configurations are also presented. The measurements are 
compared to the corresponding theoretical predictions.

\begin{figure}[b]
\includegraphics[scale=0.37,bb=0 0 550 500,clip=true]{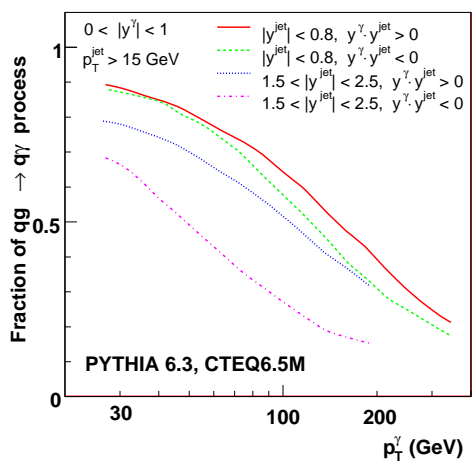}
\vspace*{-3mm}
\caption{The fraction of events, estimated using the {\sc pythia} event 
generator, produced via the $qg \to q\gamma$ subprocess relative to the total 
associated production of a direct photon and a jet for each of
the four measured configurations of the leading jet and leading photon
rapidities.}
\label{fig:subproc}
\end{figure}

Isolated final-state photons produced in \gpj ~events are expected to 
mainly originate ``directly'' from QCD Compton-like 
$qg \rightarrow q\gamma $ scattering or $q\bar{q} \rightarrow g\gamma $ annihilation subprocesses. 
In Fig.~\ref{fig:subproc} the expected contribution, estimated using 
{\sc pythia} \cite{PYT} Monte Carlo (MC) event generator  with the CTEQ6.5M parton distribution 
function (PDF) set \cite{CTEQ}, 
of the Compton-like partonic scattering process to the total associated production of a photon and a 
jet is shown for each of the four measured rapidity intervals.
The parton distribution functions entering into the theoretical predictions have substantial uncertainties, 
particularly for the gluon contributions at small $x$, large $x$ and large $Q^{2}$ \cite{WKTung,CTEQ}.
The measurement intervals probe different regions of parton momentum-fraction 
space of the two initial interacting partons, $x_{1,2}$. For example at 
$\Ptg = 40$ GeV, in events with a central leading jet, the $y^{\gamma}\!\cdot\! y^{\mathrm{jet}}>0$ region covers 
adjacent $x_1$ and  $x_2$ intervals ($0.016 \lesssim x_1 \lesssim 0.040$ and 
$0.040 \lesssim x_2 \lesssim 0.100$), while for events with $y^{\gamma}\!\cdot\! y^{\mathrm{jet}}<0$, 
the $x_1$ and  $x_2$ intervals are similar ($0.029 \lesssim x_1 \lesssim 0.074$, $0.027 \lesssim x_2 \lesssim 0.065$). 
In events with a forward leading jet, intervals of small and large $x$ are covered ($0.009 \lesssim x_1 \lesssim 0.024$, $0.110 \lesssim x_2 \lesssim 0.300$ for $y^{\gamma}\!\cdot\! y^{\mathrm{jet}}>0$ and $0.097 \lesssim x_1 \lesssim 0.264$, $0.022 \lesssim x_2 \lesssim 0.059$ for $y^{\gamma}\!\cdot\! y^{\mathrm{jet}}<0$). Here $x_{1,2}$ are defined using the leading order approximation 
$x_{1,2} = (\Ptg/\sqrt{s})(e^{\pm y^\gamma} + e^{\pm y^{\text {jet}}})$ 
\cite{PAurLindf80,JFOwens,Cont,Au2,Vo1,Mar}.

The data presented here correspond to an integrated luminosity of 
1.01~$\pm$~0.06 fb$^{-1}$ \cite{newlumi} collected using the D0 detector 
at the Fermilab Tevatron 
$p\bar{p}$ collider operating at a center-of-mass energy $\sqrt{s}=$1.96 TeV. 
A detailed description 
of the D0 detector can be found in \cite{D0_det} and only an overview of 
the detector components relevant to this analysis is given here.

\begin{figure}[t]
\includegraphics[scale=0.38,bb=10 20 500 500,clip=true]{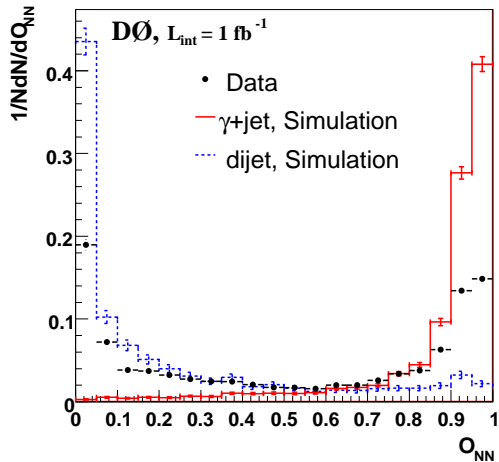}
\vspace*{-3mm}
\caption{Normalized distribution of the ANN output $O_{\rm NN}$ for data, $\gamma + \mathrm{jet}$ signal MC, and dijet background MC events 
for $44<\Ptg<50$ GeV after application of the main selection criteria. }
\label{fig:ANN_GamEmj}
\end{figure}
Photon candidates are formed from clusters of calorimeter cells in 
the central part of the liquid-argon and uranium calorimeter. 
The central calorimeter covers the pseudorapidity range 
$|\eta|<1.1$ and two end calorimeters 
cover $1.5<|\eta|<4.2$ \cite{PSR}.
The electromagnetic (EM) section of the central calorimeter contains four 
longitudinal layers of 2, 2, 7, and 10 radiation lengths, and is 
finely-segmented transversely into 
cells of size $\Delta\eta\times\Delta\phi=0.1\times0.1$  
($0.05\times0.05$ in the third EM layer), providing good angular resolution 
for photons and electrons. The position and width of the $Z$~boson mass peak,
reconstructed from $Z \rightarrow e^{+}e^{-}$ events, are used to determine 
the EM calorimeter calibration factors and the EM energy resolution 
\cite{Zgamma}. The central section of the calorimeter surrounds a central 
preshower detector, with three 
concentric cylindrical layers of scintillator strips, 
and a tracking system consisting of silicon microstrip and 
scintillating fiber trackers located within a 2~T solenoidal magnetic field.

The D0 tracking system is used to select events which contain a primary collision vertex, 
reconstructed with at least three tracks, within 50~cm of the center of the 
detector along the beam axis. The efficiency of the vertex requirement
varies as a function of instantaneous luminosity from $92\%$ to $96\%$. 

Photon candidates with rapidity $|y^{\gamma}|<1.0$ 
are selected from clusters of calorimeter cells within a cone of radius 
${\mathcal R}\equiv\sqrt{(\Delta\eta)^2+(\Delta\phi)^2}=0.4$ defined around a seed tower \cite{D0_det}.
The final cluster energy is then re-calculated from the inner cone with ${\mathcal R}=0.2$.
The data are selected using a combination of triggers based on photon EM shower profiles  
in the calorimeter and EM cluster $p_T$ thresholds. The total trigger efficiency 
is (96--97)\% for photon candidates 
with $p_T^\gamma\approx 32$~GeV and greater than $99\%$ for 
$p_T^\gamma>40$~GeV. The selected clusters are required to 
have greater than $96\%$ of their total energy contained in the EM calorimeter 
layers. Isolated clusters are selected by requiring that the energy 
$E_{\rm EM}({\mathcal R}=0.2)$, calculated within the inner cone of radius 
${\mathcal R}=0.2$, fulfills the condition 
$[E_{\rm total}({\mathcal R}=0.4)-E_{\rm EM}({\mathcal R}=0.2)]/E_{\rm EM}({\mathcal R}=0.2)<0.07$, 
where $E_{\rm total}({\mathcal R}=0.4)$ is the summed EM and hadronic energy within
a cone of radius ${\mathcal R}=0.4$.
\begin{figure}[t]
\includegraphics[scale=0.38,bb=10 20 500 500,clip=true]{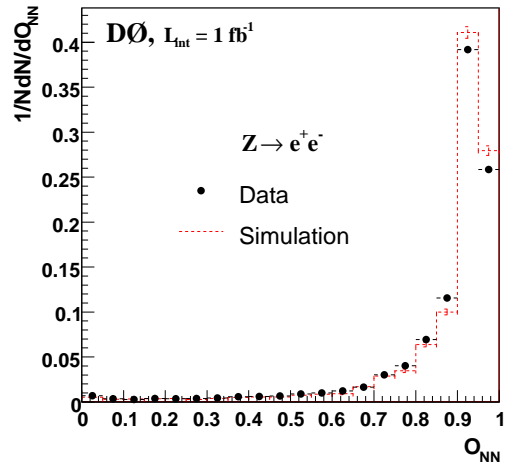}
\vspace*{-3mm}
\caption{Normalized distribution of the ANN output $O_{\rm NN}$ for electrons from $Z^0$ boson decays in data and MC events.\\ }
\label{fig:ANN_Zee}
\end{figure}
The candidate EM cluster is required not to be spatially matched to a
reconstructed track. This is accomplished by computing a $\chi^2$ function
evaluating the consistency, within uncertainties, between the reconstructed
$\eta$ and $\phi$ positions of the cluster and the closest track in the finely-segmented
third layer of the EM calorimeter. The corresponding $\chi^2$ 
probability is required to be $<0.1\%$.
\begin{figure*}
\includegraphics[scale=0.65]{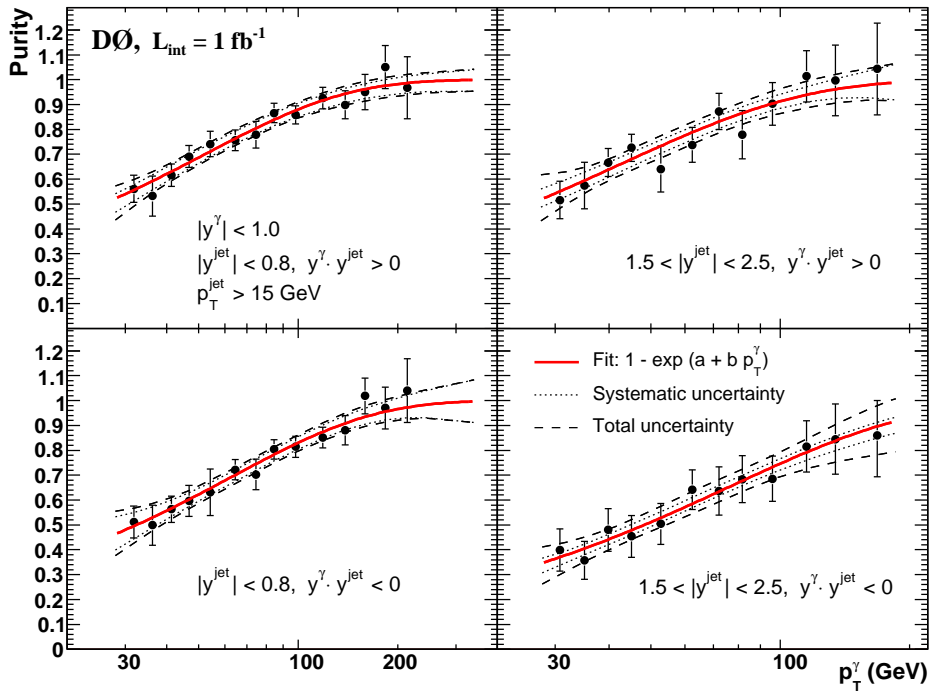} 
\vspace*{-3mm}
\caption{The purity of the selected \gpj ~sample as a function of $\Ptg$ for 
each measured configuration of photon and jet rapidities. 
The results of the 
$1 - \exp(a + bp_T^\gamma)$ functional fits are shown by the solid lines, together
with the systematic uncertainties (dotted lines), and the total uncertainties (dashed lines).}
\label{fig:pur_tot}
\end{figure*}
Background contributions to the direct photon sample from cosmic rays and 
from isolated electrons, originating from the leptonic decays of 
$W$ bosons, important at high $p_T^\gamma$ \cite{D0NoteBanSka2002},
are suppressed by requiring the missing transverse energy {\mbox{$\not\!\!E_T$}}, calculated 
as a vector sum of the transverse energies of all calorimeter cells, to satisfy
the condition ${\mbox{$\not\!\!E_T$}}<12.5 + 0.36~\Ptg$ GeV.
The longitudinal segmentation of the EM calorimeter and central preshower detector allow us to estimate
the photon candidate direction and vertex coordinate along the beam axis ("photon vertex pointing").
This vertex is required to lie within 10~cm of the event primary vertex reconstructed from charged particles.

Photons arising from decays of $\pi^0$ and $\eta$ mesons are already largely 
suppressed by the requirements above, and especially by photon isolation, 
since these mesons are produced mainly within jets during fragmentation and are surrounded by other particles.
To better select photons and estimate the residual background,
an artificial neural network (ANN) is constructed using the {\sc jetnet} package \cite{JN}. 
The following three variables are used in the ANN: the number of cells in the first EM layer belonging 
to the cluster, the fraction of the cluster energy deposited in the first
EM layer, and the scalar sum of charged particle transverse momenta in the hollow cone
$0.05 \leq {\cal R} \leq 0.4$ around the photon cluster direction. 
The resulting ANN output, $O_{\rm NN}$, after applying 
all data selection criteria, is shown, normalized to unit area, in 
Fig.~\ref{fig:ANN_GamEmj} for $44<\Ptg<50$ GeV. The output is compared to 
photon signal events and dijet background events simulated 
using {\sc pythia}.  
The signal events may contain photons originating from the parton-to-photon fragmentation process.
For this reason, the background events, produced with QCD processes in {\sc pythia}, 
were preselected to exclude the bremsstrahlung photons produced from partons. Signal and background
MC events were processed through a {\sc geant}-based \cite{Geant} simulation of the D0 detector and
the same reconstruction code as used for the data.  
The ANN is tested using 
electrons from $Z$ boson decays and the resulting normalized data and MC 
distributions are compared in Fig.~\ref{fig:ANN_Zee}. Photon candidates are 
selected by the requirement $O_{\rm NN}>0.7$ which has good background 
rejection and a signal efficiency in the range (93--97)\%. The signal selection 
efficiency decreases by about 4\% with increasing $\Ptg$ from 30 GeV to 300 GeV due to the $O_{\rm NN}>0.7$ 
requirement. The total photon+jet selection efficiency after applying all the selection 
criteria, including the ANN and the {\mbox{$\not\!\!E_T$}} requirements, is (63--77)\% as a function 
of $\Ptg$ with an overall systematic uncertainty of (4.7--5.2)\%. 
Main sources of inefficiency are the isolation, anti-track matching, 
ANN, and the photon vertex pointing cuts.

Events containing at least one hadronic jet are selected. Jets are reconstructed using the 
D0 Run II jet-finding algorithm with a cone of radius 0.7 \cite{JetAlgo}, 
and are required to satisfy 
quality criteria which suppress background from leptons, photons, and 
detector noise effects. 
Jet energies are corrected to the particle level.
The leading jet should have $p_{T}^{\mathrm{jet}}\gt 15$ GeV and $|y^{\mathrm{jet}}|\lt0.8$ or
$1.5\lt|y^{\mathrm{jet}}|\lt2.5$.
The leading photon candidate and the leading jet are also required to be 
separated in $\eta-\phi$ space by $\Delta{\cal{R}}(\gamma,\mathrm{jet})>0.7$. The  
leading jet total selection efficiency varies from 94\% to almost 100\% 
and takes into account any migrations between leading and second jet 
from the particle to the reconstruction level.
The total systematic uncertainty on this efficiency is 5.7\% at $\Ptg\simeq 30$~ GeV, 
decreasing to about 2\% at $\Ptg\ge 200$ GeV. The measurement is not very sensitive 
to jet energy scale corrections since it is performed
in bins of $\Ptg$ (with $\Ptg>30$ GeV) and only information on the the jet angular direction is used.

In total, about 1.4 million candidate events are selected after application 
of all selection criteria. A correction for the ``$\gamma$+jet'' event purity ${\cal P}$
is then applied to account for the remaining background in the region $O_{\rm NN}>0.7$.
The distribution of the ANN output for the simulated 
photon signal and dijet background samples are fitted to 
the data for each $\Ptg$ bin using a maximum likelihood fit
\cite{HMCMLL} to obtain the fractions of signal and background components in the data 
without constraining the fractions of signal and background samples in the fit to be in the $[0,1]$ range. 
The data and fitted sum of the weighted signal and background 
MC distributions of $O_{\rm NN}$ are found to be compatible with $\chi^2/ndf$ values in the 
range 0.2--1.3 \cite{Chi2Err}. The resulting purities are shown in Fig.~\ref{fig:pur_tot} for each measurement region.
The  $\Ptg$ dependence of the purity is fitted in each region using a two 
parameter function ${\cal P}=1-\exp(a+b\Ptg)$. The result of the fits 
together with their statistical errors are shown in Fig.~\ref{fig:pur_tot}.
\begin{figure}[b]
\includegraphics[scale=0.40,bb=0 0 550 380,clip=true]{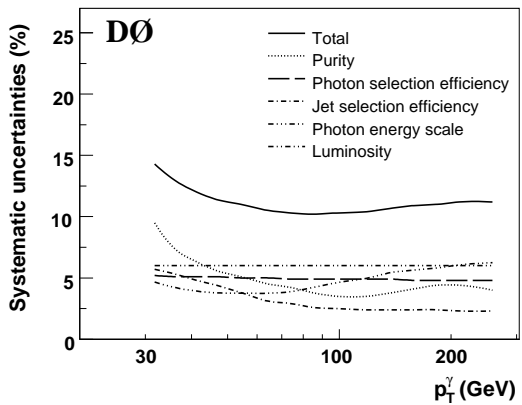}
\vspace*{-3mm}
\caption{The total and main sources of systematic uncertainty for the 
cross section measured in the $|y^{\mathrm{jet}}|\lt0.8$, $y^{\mathrm{\gamma}}\!\cdot\! y^{\mathrm{jet}}>0$ rapidity region.}
\label{fig:syst1}
\end{figure}
The systematic uncertainties on the fit are estimated using alternative 
fitting functions and varying the number of bins in the fitting
of the ANN output distribution.
An additional systematic uncertainty due to the fragmentation model 
implemented in {\sc pythia} is also taken into account.
It is found to be $5\%$ at 
$\Ptg\simeq 30$ GeV, $2\%$ at $\Ptg\simeq 50$ GeV, and $1\%$ at 
$\Ptg\gtrsim 70$ GeV \cite{Photon_paper_erratum}.

To study whether bremsstrahlung photons have different
selection efficiencies from direct photons, we extracted them from dijet events simulated with {\sc pythia}.
We found that they do not produce a noticeable change of the   
selection efficiencies, acceptance and shape of the photon ANN output that have been obtained with direct photons.

\begin{figure}[t]
\vspace*{1mm}
\hspace*{-3mm} \includegraphics[scale=0.48,bb=10 20 550 500,clip=true]{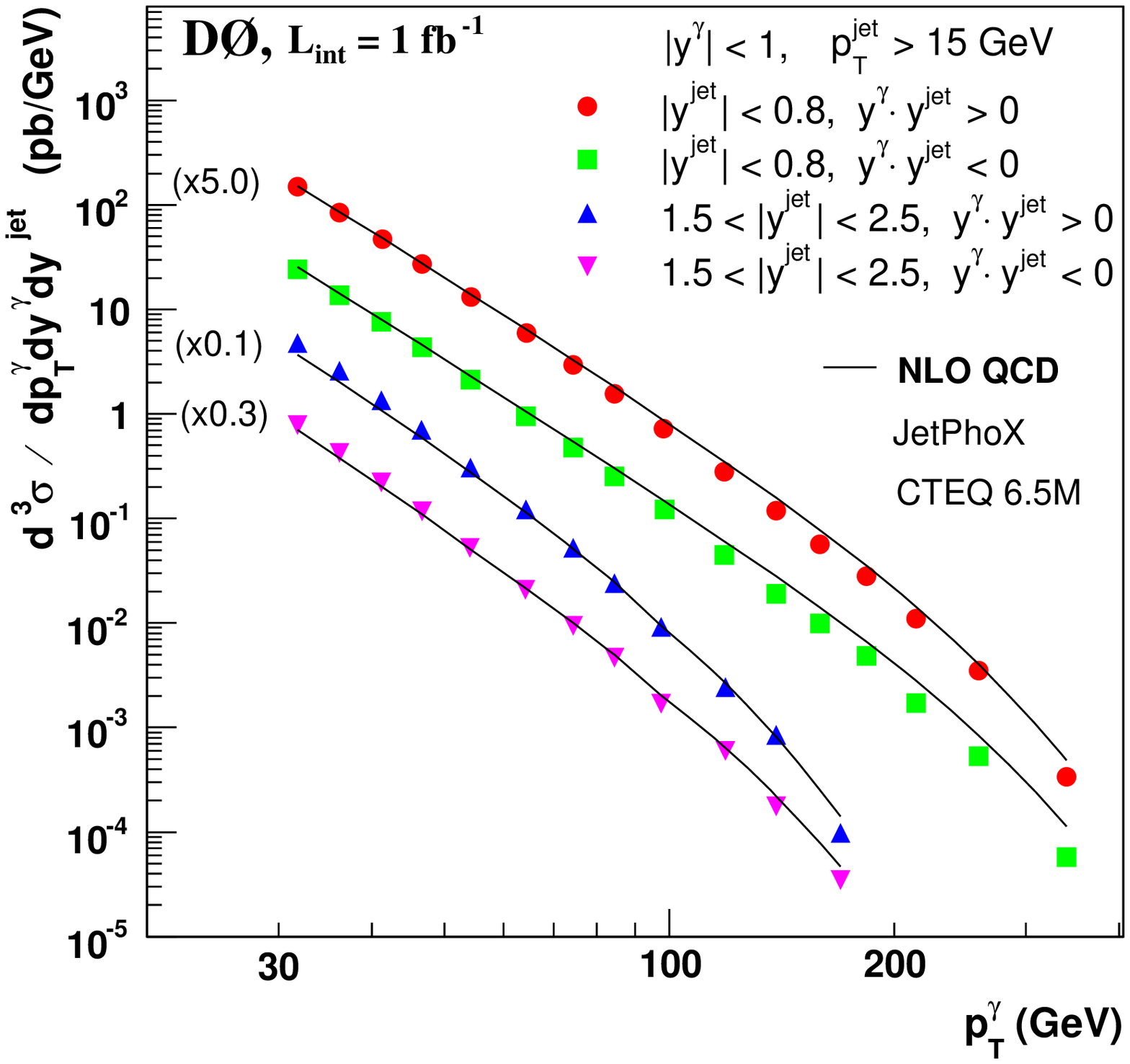}
\vspace*{-3mm}
\caption{The measured differential \gpj ~cross section as a function of
$\Ptg$ for the four measured rapidity intervals.  For presentation purposes, the cross section results for central ($|y^{\mathrm{jet}}|<0.8$) jets with $y^{\mathrm{\gamma}}\!\cdot\! y^{\mathrm{jet}}>0$ and for forward ($1.5<|y^{\mathrm{jet}}|<2.5$) jets with 
$y^{\mathrm{\gamma}}\!\cdot\! y^{\mathrm{jet}}>0$ and $y^{\mathrm{\gamma}}\!\cdot\! y^{\mathrm{jet}}<0$ are scaled by factors of 5, 0.1 and 0.3, respectively.
The data are compared to the theoretical NLO QCD predictions using the 
{\sc jetphox} package \cite{JETPHOX} with the CTEQ6.5M PDF set \cite{CTEQ} 
and renormalization, factorization and fragmentation scales $\mu_{R}=\mu_{F}=\mu_f=\newscale$.} 
\label{fig:cross}
\end{figure} 

The differential cross section $\mathrm{d^3}\sigma / \mathrm{d}\Ptg\mathrm{d}y^{\gamma}\mathrm{d}y^{\mathrm{jet}} $ for 
the process \gpj ~is obtained from the number
of data events in each interval, after applying corrections for background, efficiency, and acceptance effects, divided by the 
integrated luminosity and the widths of the interval in the photon 
transverse momentum, photon rapidity, and jet rapidity.
 The data are also 
corrected for $\Pt$ bin-migration effects which result from the finite energy 
resolution of the EM calorimeter using an analytical Ansatz 
method \cite{D0Run1_jet} and the measured EM energy resolution 
determined from the $Z$ boson peak.
The correction factors range from (1--5)\% with 
about a 1\% uncertainty.

The total ($\delta \sigma_{\rm tot}^{\rm exp}$) and main sources of experimental systematic 
uncertainty are shown for the $|y^{\mathrm{jet}}|\lt0.8$, 
$y^{\mathrm{\gamma}}\!\cdot\! y^{\mathrm{jet}}>0$ region in Fig.~\ref{fig:syst1}. 
Similar uncertainties are found for the other measured regions.
The largest uncertainties are assigned to the purity estimation [(10--4)\%], 
photon and jet selections [(7.7--5.2)\%], photon energy scale [(4.2--6.0)\%], and 
the integrated luminosity (6.1\%). 
The uncertainty ranges above are quoted with uncertainty at low $\Ptg$ first and at high $\Ptg$ second.
The systematic uncertainty on the photon selection is due mainly to the 
anti-track match cut (3\%), a correction due to observed data/MC 
difference in the efficiency of the main photon selection criteria found
from $Z\!\to\!ee$ events [(1.5--2)\%], 
the photon vertex pointing requirement (2\%), the ANN cut (2\%), and the 
uncertainty on the parameterized photon selection efficiency ($<$1\%). 
The total experimental systematic uncertainty for each data point is 
obtained by adding all the individual contributions in quadrature.

\begingroup
\squeezetable
\begin{table}
\caption{Differential cross sections $\mathrm{d^3}\sigma/\mathrm{d}\Ptg \mathrm{d}y^{\gamma}\mathrm{d}y^{\mathrm{jet}}$ and uncertainties for the $|y^{\mathrm{jet}}|\lt0.8$, $y^{\mathrm{\gamma}}\!\cdot\! y^{\mathrm{jet}}>0$ rapidity interval.}
\label{tab:cross1}               
\begin{tabular}{cccccc} \hline \hline
~~$\Ptg$ bin~  & $\la\Ptg\ra$ & \multicolumn{1}{c}{ Cross section } 
& $\delta\sigma_{\rm stat}$ & $\delta\sigma_{\rm syst}$  & $\delta\sigma_{\rm tot}^{\rm exp}$ \\
  (GeV)    & (GeV)& (pb/GeV) &  (\%) &   (\%)  &   (\%)   \\\hline
  30 --  34 &    31.9  & 3.08$\times 10^{1}$  &   0.2 &   14.2  &   14.2   \\
  34 --  39 &    36.3  & 1.74$\times 10^{1}$  &   0.3 &   13.1  &   13.1   \\
  39 --  44 &    41.3  & 9.76$\times 10^{0}$  &   0.4 &   12.4  &   12.4   \\
  44 --  50 &    46.8  & 5.60$\times 10^{0}$  &   0.5 &   11.9  &   11.9   \\
  50 --  60 &    54.6  & 2.76$\times 10^{0}$  &   0.6 &   11.5  &   11.5   \\
  60 --  70 &    64.6  & 1.24$\times 10^{0}$  &   0.9 &   11.0  &   11.0   \\
  70 --  80 &    74.7  & 6.25 $\times 10^{-1}$  &   1.2 &   10.8  &   10.9   \\
  80 --  90 &    84.7  & 3.32 $\times 10^{-1}$  &   1.7 &   10.6  &   10.7   \\
  90 -- 110 &    99.0  & 1.51 $\times 10^{-1}$  &   1.8 &   10.6  &   10.7   \\
 110 -- 130 &   119.1  & 5.79 $\times 10^{-2}$  &   2.9 &   10.5  &   10.9   \\
 130 -- 150 &   139.2  & 2.56 $\times 10^{-2}$  &   4.3 &   10.7  &   11.5   \\
 150 -- 170 &   159.3  & 1.17 $\times 10^{-2}$  &   6.5 &   10.9  &   12.7   \\
 170 -- 200 &   183.6  & 5.80 $\times 10^{-3}$  &   7.6 &   11.0  &   13.3   \\
 200 -- 230 &   213.8  & 2.33 $\times 10^{-3}$  &  11.8 &   11.0  &   16.1   \\
 230 -- 300 &   259.5  & 7.25 $\times 10^{-4}$  &  13.8 &   10.7  &   17.5   \\
 300 -- 400 &   340.5  & 7.96 $\times 10^{-5}$  &  35.3 &   10.9  &   36.9  \\\hline \hline

\end{tabular}
\end{table}
\begin{table}
\vskip -.4cm
\caption{Differential cross sections $\mathrm{d^3}\sigma/\mathrm{d}\Ptg \mathrm{d}y^{\gamma}\mathrm{d}y^{\mathrm{jet}}$  and
uncertainties for the $|y^{\mathrm{jet}}|\lt0.8$, $y^{\mathrm{\gamma}}\!\cdot\! y^{\mathrm{jet}}<0$ rapidity interval.}
\label{tab:cross2}         
\begin{tabular}{cccccc} \hline \hline 
~~$\Ptg$ bin~  & $\la\Ptg\ra$ & \multicolumn{1}{c}{ Cross section } 
& $\delta\sigma_{\rm stat}$ & $\delta\sigma_{\rm syst}$  & $\delta\sigma_{\rm tot}^{\rm exp}$ \\ 
  (GeV)    & (GeV)& (pb/GeV) &  (\%) &   (\%)  &   (\%)   \\\hline
  30 --  34 &    31.9  &  2.51$\times 10^{1}$  &   0.3 &   15.7  &   15.7   \\
  34 --  39 &    36.3  &  1.42$\times 10^{1}$  &   0.3 &   13.9  &   13.9   \\
  39 --  44 &    41.3  &  7.90$\times 10^{0}$  &   0.4 &   12.6  &   12.6   \\
  44 --  50 &    46.8  &  4.48$\times 10^{0}$  &   0.5 &   11.9  &   11.9   \\
  50 --  60 &    54.6  &  2.20$\times 10^{0}$  &   0.6 &   11.5  &   11.5   \\
  60 --  70 &    64.6  &  9.99$\times 10^{-1}$  &   0.9 &   11.1  &   11.1   \\
  70 --  80 &    74.7  &  4.98$\times 10^{-1}$  &   1.3 &   10.9  &   11.0   \\
  80 --  90 &    84.7  &  2.67$\times 10^{-1}$  &   1.8 &   10.7  &   10.9   \\
  90 -- 110 &    99.0  &  1.26$\times 10^{-1}$  &   1.9 &   10.7  &   10.9   \\
 110 -- 130 &   119.1  &  4.74$\times 10^{-2}$  &   3.1 &   10.6  &   11.1   \\
 130 -- 150 &   139.2  &  2.07$\times 10^{-2}$  &   4.7 &   10.9  &   11.9   \\
 150 -- 170 &   159.3  &  1.08$\times 10^{-2}$  &   6.6 &   11.2  &   13.0   \\
 170 -- 200 &   183.6  &  5.23$\times 10^{-3}$  &   7.7 &   11.7  &   14.0   \\
 200 -- 230 &   213.8  &  1.90$\times 10^{-3}$  &  13.0 &   11.6  &   17.4   \\
 230 -- 300 &   259.5  &  5.93$\times 10^{-4}$  &  15.0 &   11.2  &   18.7   \\
 300 -- 400 &   340.5  &  5.32$\times 10^{-5}$  &  46.1 &   12.9  &   47.8  \\\hline \hline 
\end{tabular}
\end{table}
\begin{table}
\vskip -.4cm
\caption{Differential cross sections $\mathrm{d^3}\sigma/\mathrm{d}\Ptg \mathrm{d}y^{\gamma}\mathrm{d}y^{\mathrm{jet}}$  and
uncertainties for the $1.5\lt |y^{\mathrm{jet}}|\lt2.5$, $y^{\mathrm{\gamma}}\!\cdot\! y^{\mathrm{jet}}>0$ rapidity interval.}
\label{tab:cross3}
\begin{tabular}{cccccc} \hline \hline 
~~$\Ptg$ bin~  & $\la\Ptg\ra$ & \multicolumn{1}{c}{ Cross section } 
& $\delta\sigma_{\rm stat}$ & $\delta\sigma_{\rm syst}$  & $\delta\sigma_{\rm tot}^{\rm exp}$ \\ 
  (GeV)    & (GeV)& (pb/GeV) &  (\%) &   (\%)  &   (\%)   \\\hline
  30 --  34 &    31.9  &  1.67$\times 10^{1}$  &   0.3 &   14.7  &   14.7   \\
  34 --  39 &    36.3  &  8.74$\times 10^{0}$  &   0.4 &   13.5  &   13.5   \\
  39 --  44 &    41.3  &  4.53$\times 10^{0}$  &   0.5 &   12.8  &   12.8   \\
  44 --  50 &    46.8  &  2.36$\times 10^{0}$  &   0.7 &   12.4  &   12.4   \\
  50 --  60 &    54.5  &  1.02$\times 10^{0}$  &   0.8 &   11.8  &   11.8   \\
  60 --  70 &    64.6  &  3.96$\times 10^{-1}$ &   1.4 &   11.2  &   11.3   \\
  70 --  80 &    74.6  &  1.71$\times 10^{-1}$  &   2.1 &   10.8  &   11.0   \\
  80 --  90 &    84.7  &  7.76$\times 10^{-2}$  &   3.2 &   10.8  &   11.3   \\
  90 -- 110 &    98.8  &  3.05$\times 10^{-2}$  &   3.6 &   10.7  &   11.3   \\
 110 -- 130 &   118.9  &  8.27$\times 10^{-3}$  &   6.9 &   11.0  &   13.0   \\
 130 -- 150 &   139.0  &  2.85$\times 10^{-3}$  &  11.8 &   11.5  &   16.5   \\
 150 -- 200 &   169.4  &  3.15$\times 10^{-4}$  &  23.0 &   12.1  &   26.0  \\\hline \hline 
\end{tabular}
\end{table} 
\begin{table}
\vskip -.4cm
\caption{Differential cross sections $\mathrm{d^3}\sigma/\mathrm{d}\Ptg \mathrm{d}y^{\gamma}\mathrm{d}y^{\mathrm{jet}}$ and
uncertainties for the $1.5\lt |y^{\mathrm{jet}}|\lt2.5$, $y^{\mathrm{\gamma}}\!\cdot\! y^{\mathrm{jet}}<0$ rapidity interval.}
\label{tab:cross4}              
\begin{tabular}{cccccc} \hline \hline 
~~$\Ptg$ bin~  & $\la\Ptg\ra$ & \multicolumn{1}{c}{ Cross section } 
& $\delta\sigma_{\rm stat}$ & $\delta\sigma_{\rm syst}$  & $\delta\sigma_{\rm tot}^{\rm exp}$ \\ 
  (GeV)    & (GeV)& (pb/GeV) &  (\%) &   (\%)  &   (\%)   \\\hline
  30 --  34 &    31.9  &  8.08$\times 10^{0}$  &   0.4 &   15.6  &   15.6   \\
  34 --  39 &    36.3  &  4.36$\times 10^{0}$  &   0.4 &   14.2  &   14.2   \\
  39 --  44 &    41.3  &  2.23$\times 10^{0}$  &   0.6 &   13.0  &   13.0   \\
  44 --  50 &    46.8  &  1.16$\times 10^{0}$  &   0.8 &   12.3  &   12.3   \\
  50 --  60 &    54.5  &  5.28$\times 10^{-1}$  &   1.0 &   11.7  &   11.7   \\
  60 --  70 &    64.6  &  2.08$\times 10^{-1}$  &   1.7 &   11.3  &   11.4   \\
  70 --  80 &    74.6  &  9.18$\times 10^{-2}$  &   2.6 &   11.2  &   11.5   \\
  80 --  90 &    84.7  &  4.61$\times 10^{-2}$  &   3.7 &   11.3  &   11.9   \\
  90 -- 110 &    98.8  &  1.64$\times 10^{-2}$  &   4.5 &   11.2  &   12.1   \\
 110 -- 130 &   118.9  &  5.31$\times 10^{-3}$  &   8.2 &   11.1  &   13.8   \\
 130 -- 150 &   139.0  &  1.79$\times 10^{-3}$  &  14.1 &   11.2  &   18.0   \\
 150 -- 200 &   169.4  &  3.04$\times 10^{-4}$  &  23.0 &   11.3  &   25.6  \\\hline \hline 
\end{tabular}
\end{table} 
\endgroup

The result for each region is presented as a 
function of $\Ptg$ in Fig.~\ref{fig:cross} and Tables \ref{tab:cross1}--\ref{tab:cross4}. The data 
points are plotted at the value $\langle \Ptg \rangle $ for which  a value of 
the smooth function describing the cross section 
equals the average cross section in the bin \cite{TW}.
The data cover six orders of magnitude in the cross section 
for events with
$|y^{\mathrm{jet}}|\lt0.8$, falling more rapidly over four orders of magnitude 
for events with $1.5\lt |y^{\mathrm{jet}}|\lt2.5$.

\begin{figure*}
\includegraphics[scale=0.65]{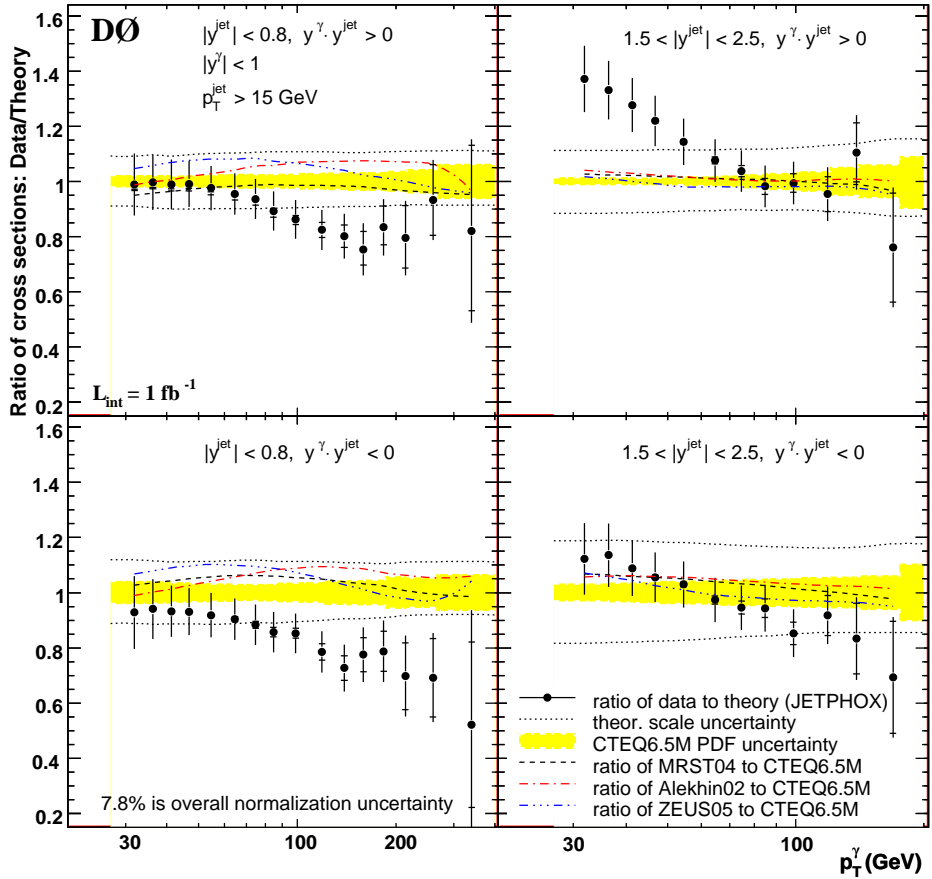} 
\caption{The ratios of the measured triple-differential cross section, in each measured interval, 
to the NLO QCD prediction using {\sc jetphox} \cite{JETPHOX}
with the CTEQ6.5M PDF set and all three scales $\mu_{R,F,f}=p_T^\gamma f({y^\star})$.
The solid vertical line on the points shows the statistical and $p_T$-dependent systematic uncertainties added in quadrature, 
while the internal line shows the statistical 
uncertainty. The two dotted lines represent the effect of varying the 
theoretical scales by a factor of two. The shaded region is the CTEQ6.5M PDF 
uncertainty. The dashed and dash-dotted lines show ratios of the {\sc jetphox} predictions with MRST~2004, 
Alekhin, and ZEUS~2005 to CTEQ6.5M PDF sets. Systematic uncertainties have large ($>80\%$) $\Ptg$ bin-to-bin
correlations.  There is a common $7.8\%$ normalization uncertainty that is not shown on the data points.
}
\label{fig:DT_reg1}
\end{figure*}

The data are compared to next-to-leading order (NLO) QCD
predictions obtained using {\sc jetphox} \cite{CFGP,JETPHOX}, with 
CTEQ6.5M PDF \cite{CTEQ} and BFG fragmentation functions of partons to photons 
\cite{BFG}. The renormalization, factorization, and 
fragmentation scales ($\mu_{R}$, $\mu_{F}$, and $\mu_f$) are set equal to
$\newscale$, where $f({y^\star})=\Fystar$ and $y^\star=\ystar$ \cite{Scale_choice}.
The theoretical predictions include selection criteria on the photon and jet
similar to those applied in the experimental analysis. In particular, an
isolation requirement on the photon of $[E_{total}({\cal R}=0.4)-E^\gamma]/E^\gamma < 0.07$
is made, where $E_{total}({\cal R}=0.4)$ is the total energy around the photon 
in a cone of radius ${\cal R}=0.4$, and  $E^\gamma$ is the photon energy. 
This requirement suppresses the relative contribution from photons produced in the fragmentation process,
and leads to a more consistent comparison with the experimental result.
Corrections for the underlying event 
and parton-to-hadron fragmentation contributions, estimated using {\sc pythia}, are found to
be negligibly small and are not included.  
To make a more detailed comparison, the ratio of the measured cross section 
to the NLO QCD prediction is taken in each interval and the results are shown 
in Fig.~\ref{fig:DT_reg1}. The inner error bars reflect the statistical 
uncertainty only, and the outer error bars are the total statistical and 
$p_T$-dependent systematic uncertainties summed in quadrature. 
Most of these systematic uncertainties, associated with the parametrizations
of the photon and jet selection efficiencies, purity (including the uncertainty from the {\sc pythia} fragmentation model),
photon $p_T$ correction, and calorimeter energy scale, have large ($>80\%$) bin-to-bin correlations in $\Ptg$.
Systematic $\Ptg$-independent uncertainties from the luminosity measurement, 
photon selection efficiency caused by the anti-track matching, ANN and photon vertex pointing, 
acceptance (1.5\%), and unfolding (1\%) 
lead to a total 7.8\% overall normalization uncertainty and are not shown in Fig.~\ref{fig:DT_reg1}. 

The prediction using the CTEQ6.5M PDF and BGF fragmentation sets does not 
describe the shape of the cross section over the whole measured range. 
In particular, the prediction is above the data for events with $|y^{\rm jet}|<0.8$ 
in the region $\Ptg \gt 100$ GeV and below the data for jets produced in the 
$1.5\lt |y^{\mathrm{jet}}| \lt 2.5$, $y^{\mathrm{\gamma}}\!\cdot\! y^{\mathrm{jet}}>0$ 
rapidity region for $\Ptg \lt 50$ GeV. 
Most of the data points in these $\Ptg$ and rapidity regions are 
(1--1.5)~$\delta \sigma_{\rm tot}$ outside of the CTEQ6.5M PDF set uncertainty range which is 
shown by the shaded region in the figure and calculated according to 
the prescription in \cite{CTEQ}. 
Note that the  data-to-theory ratios have a shape similar to those
observed in the inclusive photon cross sections measured by the UA2~\cite{UA2_phot}, 
CDF \cite{CDF_phot} and D0 \cite{Photon_paper_erratum} collaborations.

The dotted lines in Fig.~\ref{fig:DT_reg1} show the effect of setting the renormalization, 
factorization, and fragmentation scales to $0.5\newscale$ 
(upper dotted line) and $2\newscale$ (lower dotted line). The effect on
the normalization is (9--11)\%, except  for jets in the 
$1.5\lt |y^{\mathrm{jet}}|\lt 2.5$, $y^{\mathrm{\gamma}}\!\cdot\! y^{\mathrm{jet}}<0$ rapidity range 
where it is (18--20)\%. 
The scale variation is not able to simultaneously accommodate the measured differential 
cross sections in all of the measured regions. The ratios of the NLO QCD 
prediction with the  MRST~2004~\cite{MRST}, Alekhin \cite{Alekhin},  
and ZEUS~2005 \cite{ZEUS} PDF sets to the prediction obtained using the 
CTEQ6.5M PDF set are also presented in the figure. The shapes of the predictions are very similar, 
especially for forward jet production, with the different PDF sets.
 
\begin{figure}[t]
\vspace*{-0.9cm}
\includegraphics[scale=0.4, bb=0 0 550 500,clip=true]{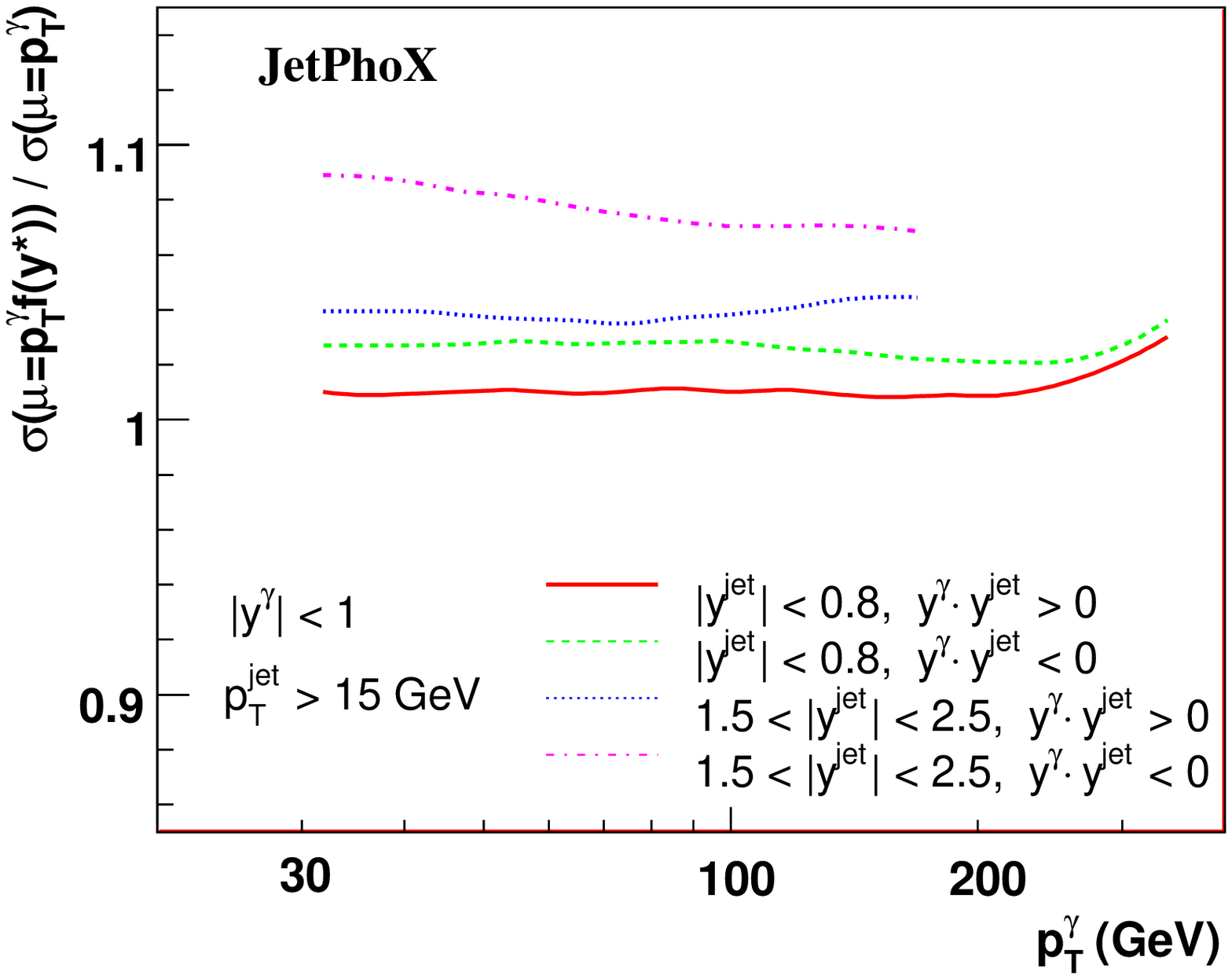}
\vspace*{-3mm}
\caption{Ratio of the predicted cross section with $\mu_{R,F,f}=\newscale$ to 
those with $\mu_{R,F,f}=\Ptg$ in each measured region.}
\label{fig:th_scale_r}
\end{figure}

The ratios of the predicted cross sections with the default scales [$\mu_{R}=\mu_{F}=\mu_f=\newscale$]
to those with all the scales 
set equal to $\Ptg$ are presented for each of the four kinematic regions as a function of $\Ptg$ in 
Fig.~\ref{fig:th_scale_r}. For each measured region, the new prediction is 
smaller than the default case across the entire $\Ptg$ range, 
most notably in the forward jet rapidity intervals 
where this choice of scale leads to a poorer level of agreement between data and theory. 

\begin{figure}[t]
\vspace*{-0.9cm} 
\includegraphics[scale=0.4,bb=0 0 550 500,clip=true]{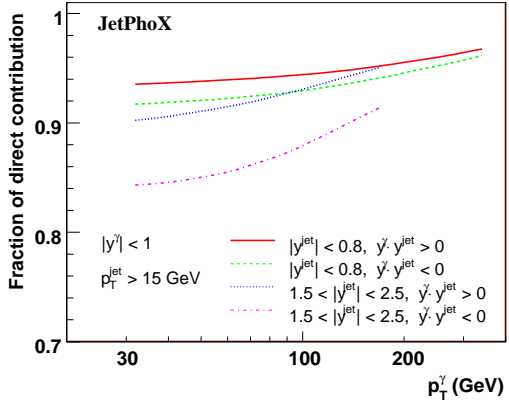}
\vspace*{-3mm}
\caption{The ratios of the \gpj ~cross section with just the direct (non-fragmentation)
contribution to the total (direct+fragmentation) cross section estimated with {\sc jetphox} for each measured 
region.}
\label{fig:dirfrac}
\end{figure}

Uncertainties related to the photon production due to the fragmentation mechanism are also studied separately 
using the {\sc jetphox} package. 
The ratio of the \gpj ~cross section for the direct photon
contribution to the sum of direct and fragmentation contributions is shown,
for the chosen photon isolation criteria, in each of the four measured 
regions in Fig.~\ref{fig:dirfrac}. For all regions, the fragmentation contribution 
decreases with increasing $\Ptg$ \cite{CFGP,Ber,PatrickPR06}
and is largest for the $1.5\lt |y^{\mathrm{jet}}| <2.5$, $y^{\mathrm{\gamma}}\!\cdot\! y^{\mathrm{jet}}<0$ region.
A variation in the fragmentation scale by a factor of 
four leads to only a (2--3)\% change in the total predicted cross section. 
Similarly a change in default set of fragmentation functions (BFG Set 1 to 
BFG Set 2) results in a cross section change of $\lesssim 1\%$. 

A possible contribution to the theoretical cross section from threshold resummation has been 
estimated \cite{Vog_resum} for inclusive direct photon production at 
the Tevatron and found to be $\lesssim (2.5-3.0)\%$ for $\Ptg \lesssim 350$ GeV.

\begin{figure*}
\includegraphics[scale=0.65]{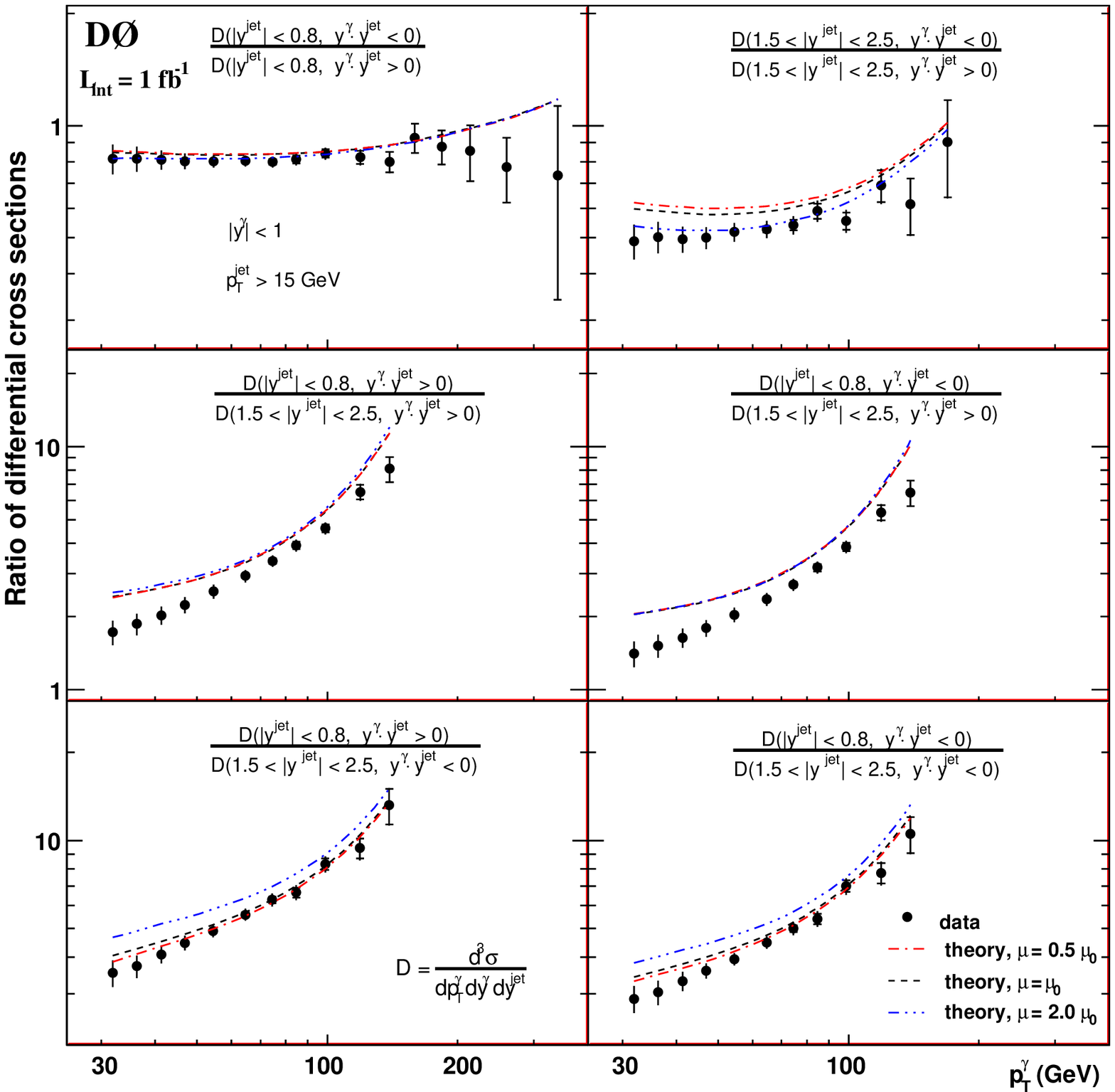}
\caption{The ratios between the differential cross sections in 
each $y^{\mathrm{jet}}$ region. The solid vertical error bars correspond to the statistical
and systematic uncertainties added in quadrature while the
horizontal marks indicate the statistical uncertainty. NLO QCD theoretical predictions 
for the ratios, estimated using {\sc jetphox}, are shown for three different 
scales: $\mu_{R,F,f}$=$\mu_{0}$, 0.5$\mu_{0}$, and 2$\mu_{0}$, where $\mu_{0}=p_T^\gamma f({y^\star})$.}
\label{fig:cross_ratio1}
\end{figure*}

The experimental systematic uncertainties are reduced further by measuring
the ratios between the differential cross sections 
$\mathrm{D}=\mathrm{d^3}\sigma / \mathrm{d}\Ptg\mathrm{d}y^{\gamma}\mathrm{d}y^{\mathrm{jet}}$ in the different regions. Most of the systematic uncertainties related to the 
identification of central photons then cancel, and only systematic 
uncertainties related to the \gpj ~event purities and the jet selection 
efficiency (when measuring ratios between central and forward jet 
regions) remain. 
Measured ratios between the differential cross sections 
in the different regions are presented in Fig.~\ref{fig:cross_ratio1} and 
Tables \ref{tab:cross_ratio21}--\ref{tab:cross_ratio24}.
The overall experimental uncertainty is largest in the first and 
last $\Ptg$ bins and ranges from (3--9)\% across most of
the $\Ptg$ range. The NLO QCD predicted cross section ratios 
estimated using {\sc jetphox} are also presented for 
scale choices $\mu_{R,F,f}=p_T^\gamma f({y^\star})$,  
$\mu_{R,F,f}=0.5 p_T^\gamma f({y^\star})$, and $\mu_{R,F,f}=2 p_T^\gamma f({y^\star})$. 
The scale uncertainty of the predicted ratios is $\leq3\%$ and about (3.5--7.5)\% 
for the ratio of cross sections in the two forward jet rapidity intervals.
The shapes of the measured ratios between the cross sections in the 
different regions, in general, are qualitatively reproduced by the
theory. A quantitative 
difference, however, between theory and the measurement is observed for the 
ratios of the central jet regions to the forward $1.5\lt |y^{\mathrm{jet}}|\lt2.5$, $y^{\mathrm{\gamma}}\!\cdot\! y^{\mathrm{jet}}>0$ 
region, even after the theoretical scale variation is taken into account.  The ratio between the 
two forward jet cross sections suggests a scale choice $\mu_{R,F,f} \simeq 2 p_T^\gamma f({y^\star})$. 
However, the ratios of the central jet regions to the forward 
$1.5\lt |y^{\mathrm{jet}}|\lt2.5$, 
 $y^{\mathrm{\gamma}}\!\cdot\! y^{\mathrm{jet}}<0$ region suggest a 
theoretical scale closer to $\mu_{R,F,f} \simeq 0.5 p_T^\gamma f({y^\star})$.


\begingroup
\begin{table}
\caption{Ratios of the differential cross sections in the 
$|y^{\mathrm{jet}}|\lt 0.8$, $y^{\mathrm{\gamma}}\!\cdot\! y^{\mathrm{jet}}<0$ 
rapidity region to the 
$|y^{\mathrm{jet}}|\lt 0.8$, $y^{\mathrm{\gamma}}\!\cdot\! y^{\mathrm{jet}}>0$ 
rapidity region.}
\label{tab:cross_ratio21}
\begin{tabular}{ccccc} \hline \hline 
~~$\Ptg$ bin~  & \multicolumn{1}{c}{Ratio } 
& ~$\delta r_{\rm stat}$~ & $\delta r_{\rm syst}$  & $\delta r_{\rm tot}^{\rm exp}$  \\
  (GeV)    &  ($r$) &  (\%) &   (\%)  &   (\%)  \\\hline
  30 --  34 &    0.81 &   0.4 &   9.1 &       9.1   \\
  34 --  39 &    0.81 &   0.4 &   7.8 &       7.8   \\
  39 --  44 &    0.81 &   0.6 &   6.0 &       6.0   \\
  44 --  50 &    0.80 &   0.7 &   4.8 &       4.8   \\
  50 --  60 &    0.80 &   0.8 &   3.7 &       3.8   \\
  60 --  70 &    0.81 &   1.3 &   3.1 &       3.3   \\
  70 --  80 &    0.80 &   1.8 &   2.9 &       3.4   \\
  80 --  90 &    0.81 &   2.5 &   2.8 &       3.8   \\
  90 -- 110 &    0.83 &   2.6 &   2.8 &       3.8   \\
 110 -- 130 &    0.82 &   4.3 &   2.7 &       5.0   \\
 130 -- 150 &    0.81 &   6.4 &   2.4 &       6.9   \\
 150 -- 170 &    0.93 &   9.3 &   2.2 &       9.5   \\
 170 -- 200 &    0.90 &  10.8 &   2.1 &      11.0   \\
 200 -- 230 &    0.81 &  17.6 &   2.0 &      17.7   \\
 230 -- 300 &    0.82 &  20.4 &   1.8 &      20.4   \\
 300 -- 400 &    0.67 &  58.0 &   0.1 &      58.0  \\\hline \hline 
\end{tabular}
\end{table} 
\begin{table}
\caption{Ratios of the differential cross sections in the 
$|y^{\mathrm{jet}}|\lt0.8$, $y^{\mathrm{\gamma}}\!\cdot\! y^{\mathrm{jet}}>0$ 
rapidity region to the \\
$1.5\lt |y^{\mathrm{jet}}|\lt2.5$, $y^{\mathrm{\gamma}}\!\cdot\! y^{\mathrm{jet}}>0$ rapidity region.}
\label{tab:cross_ratio13}
\begin{tabular}{ccccc} \hline \hline 
~~$\Ptg$ bin~  & \multicolumn{1}{c}{Ratio } 
& ~$\delta r_{\rm stat}$~ & $\delta r_{\rm syst}$  & $\delta r_{\rm tot}^{\rm exp}$  \\
  (GeV)    &  ($r$) &  (\%) &   (\%)  &   (\%)  \\\hline
  30 --  34 &    1.85  &   0.4 &  11.7 &      11.7   \\
  34 --  39 &    1.99  &   0.5 &  10.0 &      10.0   \\
  39 --  44 &    2.15  &   0.7 &   8.6 &       8.7   \\
  44 --  50 &    2.37  &   0.9 &   7.6 &       7.7   \\
  50 --  60 &    2.70  &   1.0 &   6.7 &       6.8   \\
  60 --  70 &    3.14  &   1.6 &   5.8 &       6.1   \\
  70 --  80 &    3.66  &   2.5 &   5.0 &       5.5   \\
  80 --  90 &    4.28  &   3.6 &   4.5 &       5.8   \\
  90 -- 110 &    4.97  &   4.0 &   4.2 &       5.8   \\
 110 -- 130 &    7.00  &   7.5 &   3.9 &       8.4   \\
 130 -- 150 &    9.01  &  12.6 &   3.6 &      13.1  \\\hline \hline 
\end{tabular}
\end{table} 
\endgroup

\begingroup
\begin{table}
\caption{Ratios of the differential cross sections in the 
$|y^{\mathrm{jet}}|\lt0.8$, $y^{\mathrm{\gamma}}\!\cdot\! y^{\mathrm{jet}}<0$ rapidity region to the \\
$1.5\lt |y^{\mathrm{jet}}|\lt2.5$, $y^{\mathrm{\gamma}}\!\cdot\! y^{\mathrm{jet}}>0$ rapidity region.}
\label{tab:cross_ratio23}
\begin{tabular}{ccccc} \hline \hline 
~~$\Ptg$ bin~  & \multicolumn{1}{c}{Ratio } 
& ~$\delta r_{\rm stat}$~ & $\delta r_{\rm syst}$  & $\delta r_{\rm tot}^{\rm exp}$  \\ %\cline{2-3}
  (GeV)    &  ($r$) &  (\%) &   (\%)  &   (\%)  \\\hline
  30 --  34 &    1.51 &   0.4 &  11.9 &      11.9   \\
  34 --  39 &    1.62 &   0.5 &  10.6 &      10.6   \\
  39 --  44 &    1.74 &   0.7 &   9.1 &       9.1   \\
  44 --  50 &    1.90 &   0.9 &   8.0 &       8.0   \\
  50 --  60 &    2.16 &   1.0 &   7.0 &       7.0   \\
  60 --  70 &    2.53 &   1.7 &   5.9 &       6.2   \\
  70 --  80 &    2.92 &   2.5 &   5.0 &       5.6   \\
  80 --  90 &    3.44 &   3.7 &   4.5 &       5.8   \\
  90 -- 110 &    4.14 &   4.1 &   4.3 &       5.9   \\
 110 -- 130 &    5.73 &   7.6 &   4.0 &       8.6   \\
 130 -- 150 &    7.27 &  12.8 &   3.7 &      13.3  \\\hline \hline 
\end{tabular}
\end{table} 
\begin{table}
\caption{Ratios of the differential cross sections in the 
$1.5\lt |y^{\mathrm{jet}}|\lt2.5$, $y^{\mathrm{\gamma}}\!\cdot\! y^{\mathrm{jet}}<0$ rapidity region to the \\
$1.5\lt |y^{\mathrm{jet}}|\lt2.5$, $y^{\mathrm{\gamma}}\!\cdot\! y^{\mathrm{jet}}>0$ rapidity region.}
\label{tab:cross_ratio43}              
\begin{tabular}{ccccc} \hline \hline 
~~$\Ptg$ bin~  & \multicolumn{1}{c}{Ratio } 
& ~$\delta r_{\rm stat}$~ & $\delta r_{\rm syst}$  & $\delta r_{\rm tot}^{\rm exp}$  \\ %\cline{2-3}
  (GeV)    &  ($r$) &  (\%) &   (\%)  &   (\%)  \\\hline
  30 --  34 &    0.49  &  0.5 & 10.9 &      10.9   \\
  34 --  39 &    0.50  &  0.6 &  9.7 &       9.7   \\
  39 --  44 &    0.49  &  0.9 &  8.0 &       8.0   \\
  44 --  50 &    0.49  &  1.1 &  6.7 &       6.8   \\
  50 --  60 &    0.52  &  1.3 &  5.6 &       5.8   \\
  60 --  70 &    0.53  &  2.2 &  5.0 &       5.4   \\
  70 --  80 &    0.54  &  3.4 &  4.8 &       5.8   \\
  80 --  90 &    0.59  &  4.9 &  4.8 &       6.8   \\
  90 -- 110 &    0.54  &  5.8 &  4.7 &       7.4   \\
 110 -- 130 &    0.64  & 10.8 &  4.5 &      11.7   \\
 130 -- 150 &    0.63  & 18.4 &  4.3 &      18.9   \\
 150 -- 200 &    0.97  & 32.5 &  4.1 &      32.8  \\\hline \hline 
\end{tabular}
\end{table} 
\endgroup
\begingroup
\begin{table}
\caption{Ratios of the differential cross sections in the $|y^{\mathrm{jet}}|\lt0.8$, $y^{\mathrm{\gamma}}\!\cdot\! y^{\mathrm{jet}}>0$
 rapidity region to the \\
$1.5\lt |y^{\mathrm{jet}}|\lt 2.5$, $y^{\mathrm{\gamma}}\!\cdot\! y^{\mathrm{jet}}<0$ rapidity region.}
\label{tab:cross_ratio14}              
\begin{tabular}{ccccc} \hline \hline 
~~$\Ptg$ bin~  & \multicolumn{1}{c}{Ratio } 
& ~$\delta r_{\rm stat}$~ & $\delta r_{\rm syst}$  & $\delta r_{\rm tot}^{\rm exp}$  \\ 
  (GeV)    &  ($r$) &  (\%) &   (\%)  &   (\%)  \\\hline
  30 --  34 &    3.81 &   0.4 &  10.4 &      10.4   \\
  34 --  39 &    4.00 &   0.5 &   8.8 &       8.8   \\
  39 --  44 &    4.39 &   0.8 &   6.9 &       6.9   \\
  44 --  50 &    4.82 &   1.0 &   5.5 &       5.6   \\
  50 --  60 &    5.23 &   1.2 &   4.6 &       4.7   \\
  60 --  70 &    5.97 &   1.9 &   4.3 &       4.7   \\
  70 --  80 &    6.81 &   2.8 &   4.5 &       5.3   \\
  80 --  90 &    7.20 &   4.1 &   4.6 &       6.1   \\
  90 -- 110 &    9.21 &   4.8 &   4.6 &       6.7   \\
 110 -- 130 &   10.91 &   8.7 &   4.6 &       9.9   \\
 130 -- 150 &   14.31 &  14.8 &   4.4 &      15.4   \\
 150 -- 200 &   38.29 &  23.9 &   4.2 &      24.2  \\\hline \hline 
\end{tabular}
\end{table} 
\endgroup
\begingroup
\begin{table}
\caption{Ratios of the differential cross sections in the $|y^{\mathrm{jet}}|\lt0.8$, $y^{\mathrm{\gamma}}\!\cdot\! y^{\mathrm{jet}}<0$
 rapidity region to the \\
$1.5\lt |y^{\mathrm{jet}}|\lt 2.5$, $y^{\mathrm{\gamma}}\!\cdot\! y^{\mathrm{jet}}<0$ rapidity region.}
\label{tab:cross_ratio24}              
\begin{tabular}{ccccc} \hline\hline 
~~$\Ptg$ bin~  & \multicolumn{1}{c}{Ratio } 
& ~$\delta r_{\rm stat}$~ & $\delta r_{\rm syst}$  & $\delta r_{\rm tot}^{\rm exp}$  \\ 
  (GeV)    &  ($r$) &  (\%) &   (\%)  &   (\%)  \\\hline
  30 --  34 &    3.10  &   0.4 &  10.7 &      10.7   \\
  34 --  39 &    3.25  &   0.5 &   9.5 &       9.5   \\
  39 --  44 &    3.55  &   0.8 &   7.4 &       7.5   \\
  44 --  50 &    3.86  &   1.0 &   6.0 &       6.1   \\
  50 --  60 &    4.18  &   1.2 &   4.9 &       5.0   \\
  60 --  70 &    4.81  &   1.9 &   4.5 &       4.9   \\
  70 --  80 &    5.43  &   2.9 &   4.5 &       5.4   \\
  80 --  90 &    5.80  &   4.1 &   4.6 &       6.2   \\
  90 -- 110 &    7.67  &   4.9 &   4.7 &       6.8   \\
 110 -- 130 &    8.93  &   8.8 &   4.7 &      10.0   \\
 130 -- 150 &   11.55  &  14.9 &   4.5 &      15.5   \\
 150 -- 200 &   35.48  &  23.9 &   4.3 &      24.3 \\\hline\hline 
\end{tabular}
\end{table} 
\endgroup

In summary, the differential cross section 
$\mathrm{d^3}\sigma / \mathrm{d}\Ptg\mathrm{d}y^{\gamma}\mathrm{d}y^{\mathrm{jet}} $ for 
the process $p\bar{p}\rightarrow \gamma + \mathrm{jet} + X$ is measured for  
central photons ($|y^\gamma|\lt 1.0$) separately for four different rapidity
configurations between the leading photon and the leading jet. The data 
cover six orders of magnitude in the cross section as a function of $\Ptg$ for events with jets in $|y^{\rm jet}|<0.8$,  
and extend the kinematic reach of previous photon plus jet measurements. Next-to-leading order QCD 
predictions, using a few different modern parameterizations of parton distribution 
functions, are unable to describe the shape of the $\Ptg$ dependence 
of the cross section across the entire measured range. Similarly, theoretical
scale variations are unable to simultaneously describe
the data-to-theory ratios in each of the four measured regions.
Thus, the data presented in this Letter,
show a need for an improved and consistent theoretical description of 
the $\gamma+$jet production process. 

We are very thankful to P.~Aurenche, M.~Fontannaz, J. P.~Guillet, and M.~Werlen 
for providing the {\sc jetphox} package, useful discussions and assistance 
with theoretical calculations.
%
% acknowledgement_paragraph_r2.tex                                 8/20/07
%
We thank the staffs at Fermilab and collaborating institutions, 
and acknowledge support from the 
DOE and NSF (USA);
CEA and CNRS/IN2P3 (France);
FASI, Rosatom and RFBR (Russia);
CAPES, CNPq, FAPERJ, FAPESP and FUNDUNESP (Brazil);
DAE and DST (India);
Colciencias (Colombia);
CONACyT (Mexico);
KRF and KOSEF (Korea);
CONICET and UBACyT (Argentina);
FOM (The Netherlands);
Science and Technology Facilities Council (United Kingdom);
MSMT and GACR (Czech Republic);
CRC Program, CFI, NSERC and WestGrid Project (Canada);
BMBF and DFG (Germany);
SFI (Ireland);
The Swedish Research Council (Sweden);
CAS and CNSF (China);
Alexander von Humboldt Foundation;
and the Marie Curie Program.
%%%% remove Marie Curie at August 07 update
%

%
\bibliography{paper_final}

\begin{thebibliography}{33}
\bibitem{RAP}  
\bibfnamefont{$y = 1/2 ln(E-p_L)/(E+p_L)$, where $E$ is the energy and
$p_L$ is the longitudinal momenta with respect to the $z$ axis.}
\bibitem{PSR} 
\bibfnamefont{Pseudorapidity is defined as $\eta=-\ln[\tan(\theta/2)]$, where
$\theta$ is the polar angle with respect to the proton beam direction.}
\bibitem{Chi2Err}
\bibfnamefont{Only statistical uncertainties in the $\gamma$+jet MC, dijet MC and data samples are taken into account
in the calculation of $\chi^2$}.
\bibitem{Uncert}
\bibfnamefont{The uncertainty ranges are quoted for a low $\Ptg$ first and for a high $\Ptg$ second.}
\end{thebibliography}
\bibliographystyle{apsrev}

\end{document}